\documentclass[a4paper,12pt]{article} 
\usepackage{graphicx}
\usepackage{amssymb}
\usepackage{amsmath}
\usepackage[colorlinks=true, pdfstartview=FitV, linkcolor=blue,  citecolor=blue, urlcolor=blue]{hyperref}  %liens hypertexte
\usepackage{float}
\usepackage{yfonts}
\usepackage[T1]{fontenc}

\graphicspath{ {images/} }

\newcommand{\Ds}{{\widetilde D}} % D rond
\def\l{\left}
\def\r{\right}
\def\beq{\begin{equation}}
\def\eeq{\end{equation}}
\def\d{\partial}

\def\beq{\begin{equation}}\def\eeq{\end{equation}}
\def\bea{\begin{eqnarray}}\def\eea{\end{eqnarray}}

\usepackage{parskip}

\begin{document}

\title{A new {\it{ab initio}} approach to the development of high temperature super conducting materials.}

\author{Philip Turner$^1$\footnote{ph.turner@napier.ac.uk}  and Laurent Nottale$^2$\footnote{laurent.nottale@obspm.fr}\\
$^1${\small Edinburgh Napier University, 10 Colinton Road, Edinburgh, EH10 5DT, United Kingdom.} \\
$^2${\small CNRS, LUTH, Observatoire de Paris-Meudon, 5 Place Janssen, 92190, Meudon, France.}} 

\maketitle
\begin{abstract}

We review recent theoretical developments, which suggest that a set of shared principles underpin macroscopic quantum phenomena observed in high temperature superconducting materials, room temperature coherence in photosynthetic processes and the emergence of long range order in biological structures.  These systems are driven by dissipative systems, which  lead to fractal assembly and a fractal network of charges (with associated quantum potentials) at the molecular scale.  At critical levels of charge density and fractal dimension, individual quantum potentials merge to form a `charged-induced' macroscopic quantum potential, which act as a structuring force dictating long-range order.  Whilst the system is only partially coherent (i.e. only the bosonic fields are coherent), within these processes many of the phenomena associated with standard quantum theory are recovered, with macroscopic quantum potentials and associated forces having their equivalence in standard quantum mechanics.

 We establish a testable hypothesis that the development of structures analogous to those found in biological systems, which exhibit macroscopic quantum properties, should lead to increased critical temperatures in high temperature superconducting materials.  If the theory is confirmed it opens up a new, systematic, {\it{ab initio}} approach to the structural development of these types of materials.

\end{abstract}

\section*{\centering{1. Introduction}}

As a field of research, the understanding of quantum systems in complex matter is a growing area of interest.  One of the most active areas has been associated with high temperature super conducting (HTSC) materials, leading to the development of a range of new materials exhibiting incremental increases in critical temperature $T_c$.  However, despite significant progress, there is still lack of consensus on the origins of macroscopic quantum coherence with a number of open questions to be resolved at a fundamental level, which would allow a systematic, {\it{ab initio}} approach to their on-going development and the ultimate goal of room temperature superconductivity. 

In parallel with developments in HTSC materials there has been considerable progress within the biophysics community, with confirmation of macroscopic quantum coherence in complex protein systems associated with photosynthetic systems, which operate at ambient temperatures \cite{Engel2007,Collini2010,Romero2014}.  These developments are supported by recent theoretical work \cite{Vattay2015}, suggesting that protein structures have evolved as a key component of biological systems, precisely because their complex structures exhibit macroscopic quantum properties, which play a key role in biochemical electronic processes.  This combination of theory and experiment is backed up by new, more general theory \cite{Turner2016}, which in addition, suggests that quantum criticality in complex matter also plays a fundamental role in the emergence of long range order in living systems.

In this paper we review progress in these fields, outlining a common set of theoretical principles that we suggest could assist in the on-going development of HTSC materials using an approach, which builds on our understanding of the emergence of quantum critical structures in biological systems.

\section*{{2. The origin of macroscopic quantum coherence in HTSC.}}

\begin{figure}[!ht]
\begin{center}
\includegraphics[width=10cm]{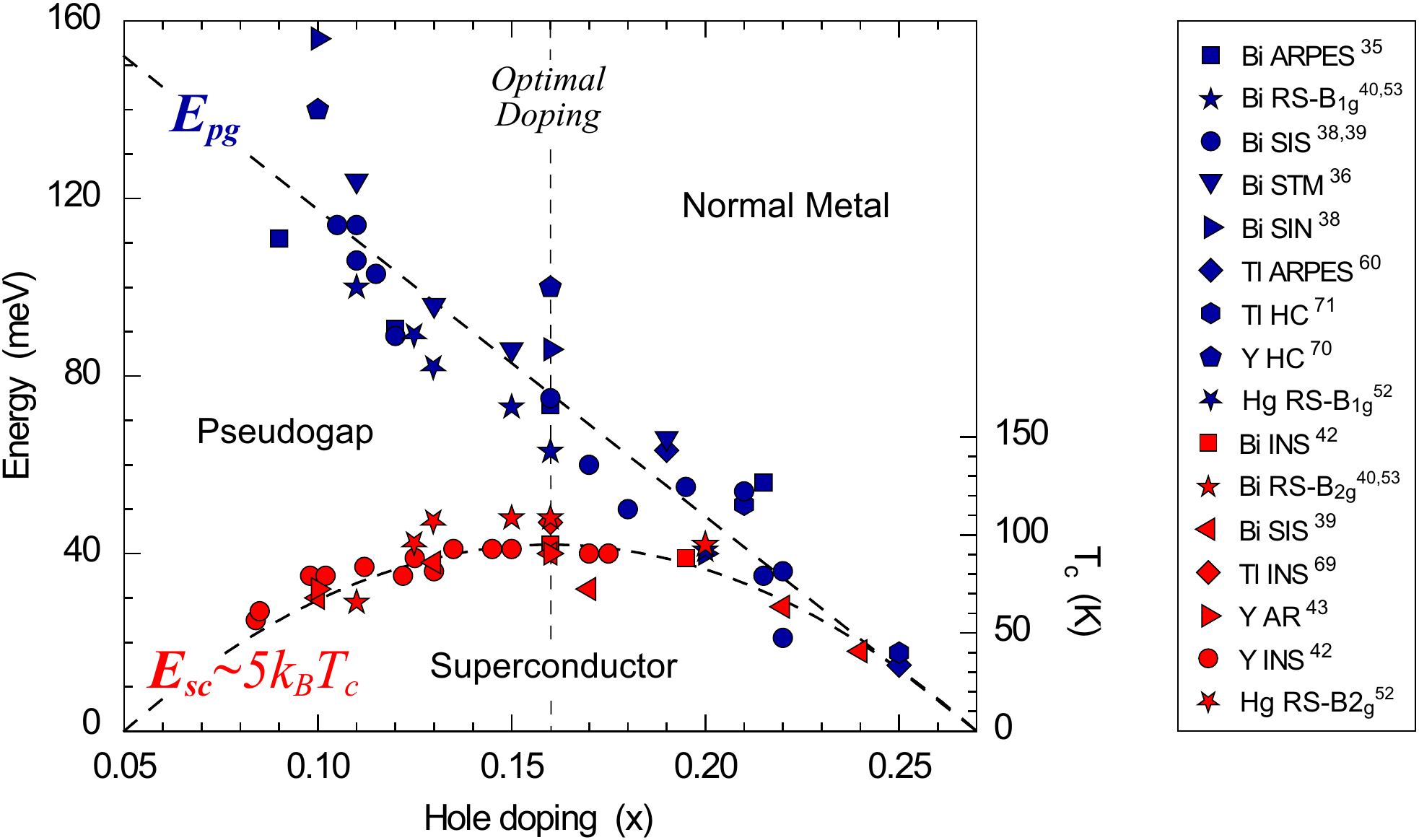} %
\caption{\small{Phase diagram for the pseudogap and superconducting gap.}}
\label{SCdome}
\end{center}
\end{figure}

In 2015, we proposed a new theoretical approach to explain relations between critical temperature of the superconducting gap $T_c$ and the Pseudogap $T^*$ observed in the $p$-type cuprates \cite{Turner2015}, an example of which is shown in Fig~\ref{SCdome}, reproduced from H\"ufner {\it{et al}} \cite{Hufner2008}.  A key element of the theory was based on findings \cite{Fratini2010} indicating a heterogeneous, scale free (fractal) distribution of static dopants $\psi_d$, whose correlation length, increases with dopant (charge) density $\rho$ up to optimum doping and then declines as $\rho$ continues to increase (associated with a reduction in correlation length), as space constraints lead to dopant packing into a more homogenous lattice arrangement \cite{Turner2015}.  

\begin{figure}[!ht]
\begin{center}
\includegraphics[width=12cm]{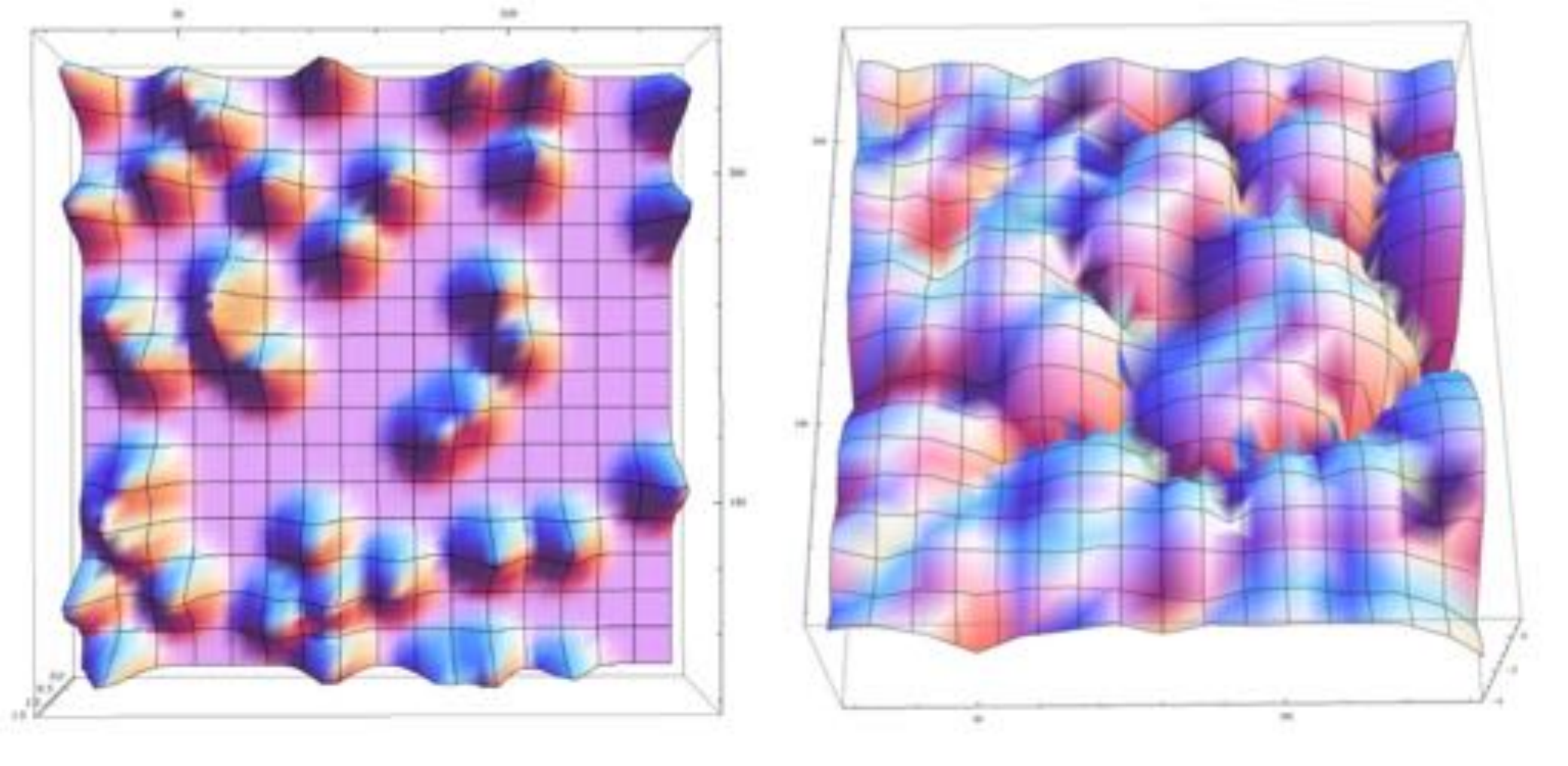} %
\caption{\small{Individual charges (left) collectively induce a fractal MQP (right.)}}
\label{charges}
\end{center}
\end{figure}

Against this background, we developed theory \cite{Turner2015} that $T_c$ relations are linked with the emergence of the fractal network of charge induced quantum potentials which at a critical percolation threshold, merge, leading to a transition from a collection of charges $\psi_d = \sqrt{\rho_d} \times e^{{i A_d}/\hbar}$ (where $A_d$ is a microscopic action), into macroscopic fluctuations, creating a complex geodesic velocity field (Fig~\ref{charges})\footnote{which can be considered as a complex macroscopic path integral \cite{Feynman1965}}, represented by a macroscopic wave function $\psi_D$

\beq
\displaystyle\sum\limits_{n=1}^N\psi_{d} \to  \psi_{D} = \sqrt{\rho_{D}}\times e^{i A_D/2\Ds},
\label{eq.68}
\eeq
where $A_D$ is a macroscopic action, $Q_D$ (Eq \ref{QMG}) is $\psi_{D}$'s associated macroscopic quantum potential (MQP) and  $\hbar$ is substituted with a macroscopic parameter $\Ds$, which characterizes the amplitude of fractal fluctuations across the MQP and is therefore specific to the system. We recall that in the case of standard QM, $\hbar$; which is itself a geometric property of a fractal space, is defined through the fractal fluctuations as $\hbar = 2m\Ds$ \cite{Turner2016,Turner2015}.    

Taking $\psi_{D}$ and $Q_D$ we write a macroscopic Schr\"odinger-like equation (Eq \ref{Schrodinger})\footnote{or non linear equation such as a Ginzburg-Landau-like equation \cite{Turner2015}}, or its equivalent as  Euler and continuity equations (Eq's \ref{Euler} and \ref{cont}), in which $Q_D$ (Eq \ref{QMG}) becomes explicit \cite{Turner2015}.

\beq
Q_{{D}}  = -2{\Ds}^2 \, \frac{\Delta \sqrt{\rho_{D}}}{\sqrt{\rho_{D}}}.
\label{QMG}.
\eeq

\beq
{\Ds}^2 \Delta \psi_D + i {\Ds} \frac{\partial\psi_D}{\partial t} - \l(\frac{\phi}{2}\r)\psi_D = 0
\label{Schrodinger}.
\eeq

\beq
\frac{\d V_{D}}{\d t} + V_{D}. \nabla V_{D} =- \frac{\nabla \phi}{m} -\frac{\nabla Q_D}{m},
\label{Euler}
\eeq
\beq
\frac{\d \rho_D }{ \d t} + \text{div}(\rho_D V_D)=0,
\label{cont}
\eeq

As a next step we consider diffusive and quantum systems as a pair of Euler equations, whose detailed derivation and full meaning is outlined in previous papers  \cite{Turner2016,Turner2015,Nottale2009}.  Eq \ref{quant} demonstrates a clear equivalence between a standard fluid subjected to a force field and a diffusion system Eq. \ref{DP}, where $D$ represents a standard diffusion coefficient, with the force expressed in terms of probability density $P$ at each point and instant.  
\begin{equation}
\l(\frac{\partial}{\partial t} + V.\nabla\r) V  = +2{\Ds}^2 \, \nabla \l(\frac{\Delta \sqrt{P}}{\sqrt{P}}\r).
\label{quant}
\end{equation}

\beq
\l( \frac{\d}{\d t} + V. \nabla\r)V=- 2 D^2 \, \nabla \l( \frac{\Delta \sqrt{P}}{\sqrt{P}}\r).
\label{DP}
\eeq

The `diffusion force' derives from an external potential
\beq
\phi_{\rm diff}=+2D^2 {\Delta \sqrt{P}}/{\sqrt{P}},
\label{diffpot}
\eeq			
which introduces a square root of probability in the description of a classical diffusion process.  The quantum force is the exact opposite, derived from the 'quantum potential', which is internally generated by its fractal geodesics 
\beq
Q/m=-2{\Ds}^2 {\Delta \sqrt{P}}/{\sqrt{P}}.
\label{quantpot}
\eeq	

Eq's \ref{quant} and \ref{DP} offer a fundamental new insight into quantum decoherence in both standard and macroscopic quantum systems \cite{Turner2016}.  The two forms of potential energy compete in a process described by the model of `quantum Brownian motion' \cite{Zurek2003,Schlosshauer2014}.  The emergence of `pointer states' during the decoherence process is linked to the collapse of a fundamental fractal root structure (underpinning the fractal velocity field) to its more stable roots \cite{Turner2016}.  These roots form the preferred set of states of an open system most robust against environmental interaction \cite {Joos2003,Kubler1973,Paz1993,Zurek1993,Diosi2000,Zurek2013}, accounting for the transition from a probabilistic to a deterministic classical description \cite{Turner2016}.  

Within the context of the $p$-type cuprates, the charge induced fractal landscape of hills and valleys forming the MQP ($Q_D$), visualized in Fig \ref{charges}, plays a key role in HTSC, providing both a macroscopic quantum force supporting macroscopic coherence, and (as supported by a number of papers \cite{Fratini2010,Poccia2011,Poccia2011b,Poccia2012}) a quantum critical, scale free network of channels for the observed superconducting quantum fluid $\psi_C$ reported by McElroy {\it{et al}} \cite{McElroy2005}.  As can be seen from Fig \ref{charges}, geometric constraints associated with the system mean that $\psi_C$ is inevitably anticorrelated with dopant induced potentials \cite{Turner2015}, in a mechanism which also explains the observed anticorrelation between dopant distribution and charge density wave order recently confirmed by Campi {\it{et al}} \cite{Campi2015}.

The outlined theory indicates that the frequency and extent of fluctuations of the MQP is dependent on the fractal dimension $D_F$ and correlation length of the scale free network. Depending on the base material, the conducting fluid may be comprised of electron-pairs coupled by one or more coupling mechanisms \cite{Turner2015}, with electron-pair (e-pair) coupling energies significantly higher than conventional SC, leading to more thermally stable e-pairs.  

Following a process analogous to the dopant system described in Eq's \ref{eq.68}-\ref{cont}, these coherent fluids ($\psi_{C}$) can also be written as a set of Schr\"odinger, Euler and continuity equations \cite{Turner2015}.  However, we are now dealing with a `multiple component' system, where the full set of equations (see Eq's 85-92 \cite{Turner2015}) includes $Q_D$ Eq (\ref{QMG}) as an external potential in the Schr\"odinger equation.  This system of disordered dopants and multiple condensates, which leads to macroscopic quantum coherence is alternatively referred to as `superstripes' \cite{Bianconi2000}; the symmetry breaking associated with quantum criticality of the fractal network leading to a classical to quantum `Lifschitz' transition and an associated `Feshbach' resonance.

One of the key limitations associated with the $p$-type cuprates lies in their 2 dimensional lattice structure  \cite{Turner2015}.  It has been proposed \cite{Turner2015} that if a continuous (scale free) 3D macroscopic complex path integral (built from the molecular scale upwards) could be developed, then significant increases in charge density and MQP strength (Eq \ref{QMG}), combined with increase in e-pair density of states and collective correlation energy, should theoretically lead to a significant increase in $T_c$.  This hypothesis is supported in principle by the observation of room temperature macroscopic quantum coherence linked to charged, fractal macromolecular systems (proteins) associated with photosynthesis \cite{Engel2007,Collini2010,Romero2014,Vattay2014}.

\subsection*{\centering{3. An {\it{\textbf{ab initio}}} approach to the development of materials supporting macroscopic quantum coherence}}

Having considered the role of fractal networks and charge density in the emergence of macroscopic quantum coherence in the $p$-type cuprates, we reflect on the development of 3D fractal structures identified with quantum criticality \cite{Vattay2015}, that could theoretically lead to enhanced $T_c$ in HTSC materials.  Recent work \cite{Turner2016,Norrduin2013} reports on the spontaneous emergence of a wide range of plant-like structures from inorganic matter, which supports theory that long range order in plant structures is driven by macroscopic quantum mechanics-type processes underpinned by a fractal network of charge density $\rho$ \cite{Turner2016}.  

The assembly of molecular-scale fractal networks is driven by a combination of quantum vacuum and thermal fluctuations acting as a sea of harmonic oscillators in combination with charge induced repulsive forces between adjacent charged particles, which influence the dynamics of molecular assembly \cite{Turner2016}. In these systems, charge density links to the solvation of $CO_2$, which leads to the release of protons and the subsequent ionization of molecules, with $\rho$ determined by $CO_2$ concentration. Increasing $\rho$ leads to an increase in the degree of molecular freedom to interact with environmental fluctuations \cite{Turner2016}.  

Successive solutions during time evolution of the time-dependent Schr\"odinger equation in a 3D harmonic oscillator potential lead to a model of branching/bifurcation and the emergence of a 3D fractal architecture, whose fractal dimension $D_F$ is dependent on $\rho$ and $T$, for a specific molecular composition \cite{Turner2016}.  During this process, charge induced potential well's will interconnect, creating a 3D version of Fig~\ref{charges}, i.e., a charge induced complex fractal velocity field.

As in Eq \ref{eq.68}, this has the effect of transforming quanta of fractal fluctuations into macroscopic fluctuations, represented by a macroscopic wave function which, with its associated MQP (Eq \ref{QMG}), can be incorporated into a set of Euler, continuity and Schr\"odinger-like equations, equivalent to those described for HTSC materials in Eq's \ref{Schrodinger} - \ref{cont} \cite{Turner2016}.

Within such a system, the MQP creates a `structuring' macroscopic quantum-like force dictating long range molecular assembly, in competition with an external diffusive force associated with the diffusive potential (Eq \ref{diffpot}) which, depending on conditions, leads to the emergence of the wide range of structures observed in the plant kingdom \cite{Turner2016}.  An example of the principle is given in Figure 3, which shows the generation of a flower-like structure, modelled using a Schr\"odinger equation \cite{Turner2016,Nottale2008}.

\begin{figure}[!ht]
\begin{center}
\includegraphics[width=10cm]{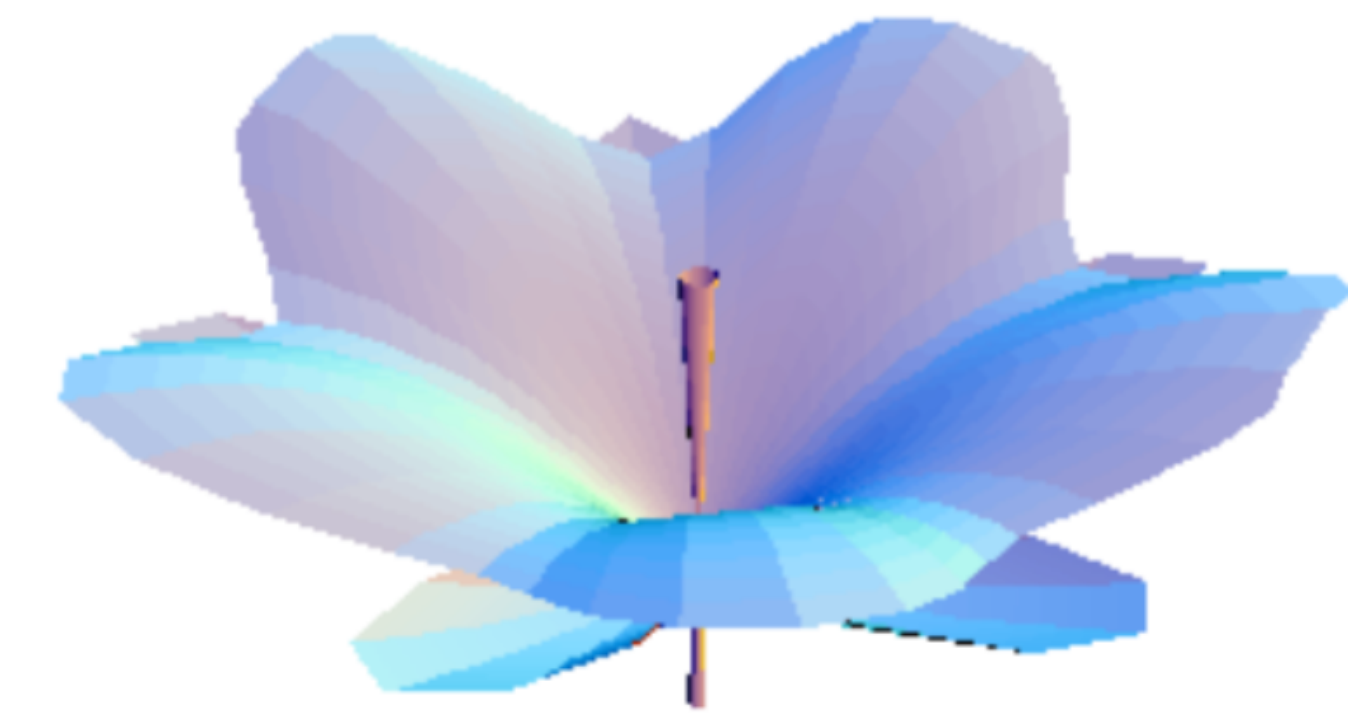} %
\caption{\small{ A flower-like structure, reproduced from Nottale and Auffray \cite{Nottale2008}.  The flower is a solution of a Schr\"odinger equation describing a growth process from a centre, with `petals', `sepals' and `stamen' traced along angles of maximal probability density.  }}
\label{Flower}
\end{center}
\end{figure}

\begin{figure}[!ht]
\begin{center}
\includegraphics[width=16cm]{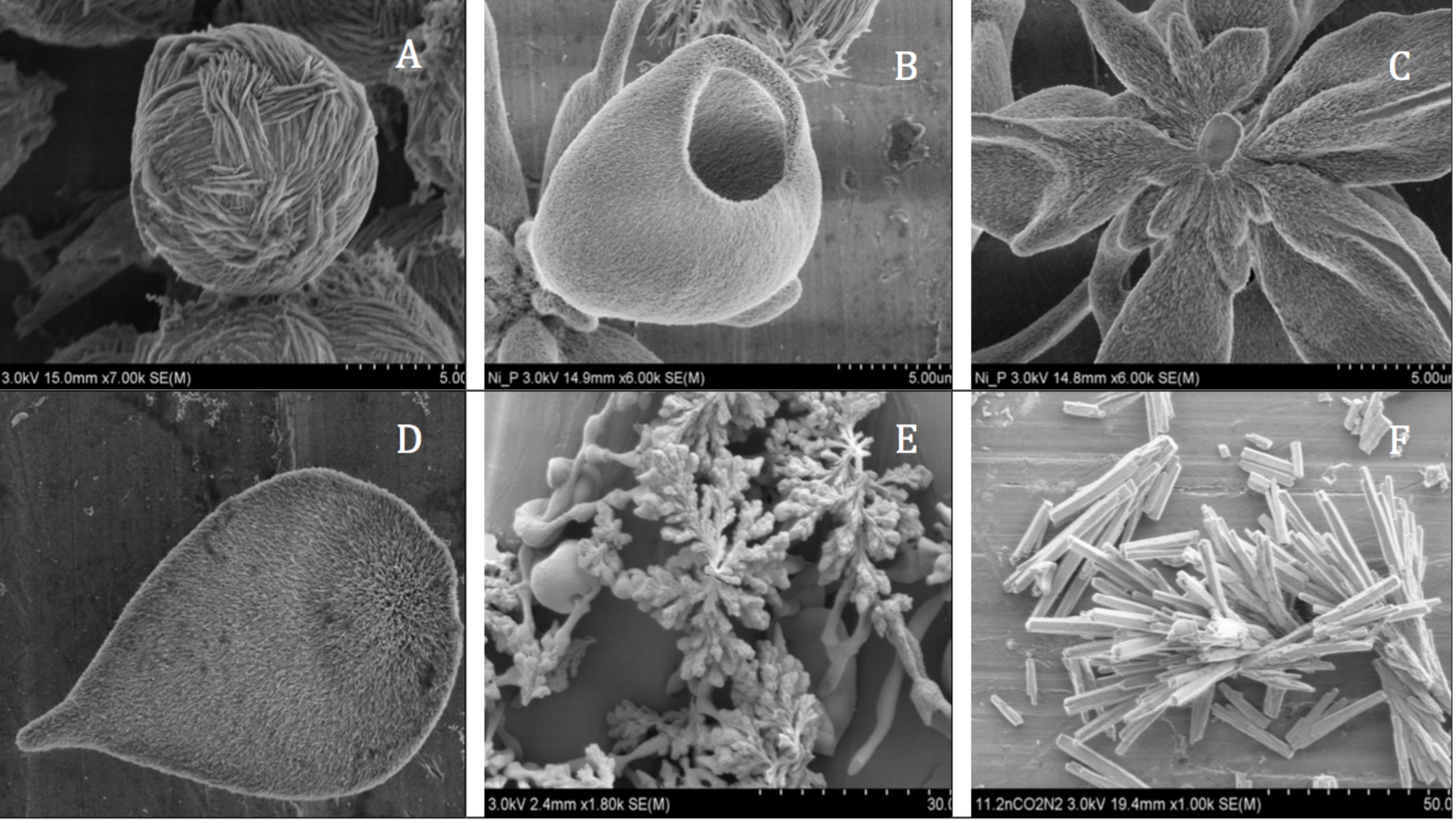} %
\caption{\small{ Emergent structures dependent on charge density showing spherical (A), pod (B), flower (C), Leaf (E), dendritic (E) and crystalline (F) structures.}}
\label{Structures}
\end{center}
\end{figure}

Considering experimental results, Fig~\ref{Structures} illustrates six examples (electron micrographs) of $BaCO_3-SiO_2$ based plant-like structures reproduced from Turner and Nottale \cite{Turner2016} at varying levels of $\rho$.  Above a critical level of $\rho$, structure takes the form of a confined $8\mu$m diameter, spherical cell-like structure Fig~\ref{Structures}A.  For a fixed temperature, a decline in $\rho$ leads to a subsequent decline in the strength of the MQP in a process of macroscopic decoherence of the complex velocity field to its fractal roots (pointer states).  This process, expressed through a range of intermediate structures of declining spatial coherence such as pod, flower and leaf-like structures in Fig's~\ref{Structures}B-D, eventually leads to structures with long-range fractal order (Fig~\ref{Structures}E). When $CO_2$ is eliminated completely from the system by purging with nitrogen, $ \rho\rightarrow 0$, the MQP collapses completely, with molecules, unhindered by repulsive charges forming a crystal lattice (Fig~\ref{Structures}F).

Apart from monocultures of spherical or crystalline structure (Figs~\ref{Structures}A and \ref{Structures}F), at ambient levels of $CO_2$, precise control over intermediate structures found in plants was unachievable, with a range of different structures emerging in response to subtle variation in local conditions at the point of growth.  This indicates that in real plants, local conditions at the point of molecular assembly during growth processes must be more tightly controlled.  An example to illustrate this \cite{Turner2016} is seen in the deposition of cellulose in secondary wall thickening in tracheary elements of {\it{Arabidopsis thaliana}} reported by Derbyshire {\it{et al}} \cite{Derbyshire2015}.  In this study involving individual stem cells in vitro, it was shown that microtubules provide the guide for cellulose synthase complexes in combination with a large number (605) of different microtubule-associated proteins, which form complexes that vary in composition as cell wall structures change during cellulose deposition.

During the transcription of DNA through RNA to an array of different proteins (which are a direct reflection of the genetic code), the subsequent assembly of protein complexes and their charge density is precisely controlled \cite{Derbyshire2015}, offering fine levels of control over average charge density on a protein complex.  As specific protein complexes attach to a growing microtubule, they dictate not only microtubule structure, but also cellulose assembly, through interaction of local levels of charge density with the cellulose synthase complex.  This leads to levels of structural control unachievable in the experimental work on inorganic plant-like structures \cite{Turner2016}.

The complexity associated with control of just one small detail in cell wall assembly in {\it{Arabidopsis thaliana}} highlights the enormous challenges associated with understanding the emergence of structure in plants and living systems in general.  However, whilst a full understanding of even simple biological systems may be decades away, the basic principles identified offer new fundamental insight into the interplay between charge density, thermodynamics and macroscopic quantum processes and their role in defining long range order in biological systems.

The macroscopic quantum processes we have described in plants take the form of a two-component [coherent (boson)-classical (fermion)] system.  Molecular assembly operates within the framework of macroscopic fluctuations within a charge induced coherent bosonic field, acting as a structuring force in competition with exterior potentials.  Whilst the system is only partially coherent, many of the phenomena associated with standard quantum theory are recovered, including quantization, non-dissipation, self-organization, confinement, structuration conditioned by the environment, environmental fluctuations leading to macroscopic quantum decoherence and evolutionary time described by the time dependent Schr\"odinger equation, which describes models of bifurcation an cell duplication \cite{Turner2016}. 

These principles offer fresh insight into mechanisms to control molecular assembly in the development of new HTSC materials.  If theory is confirmed, it offers the potential to take their macroscopic quantum properties to new levels exemplified by protein complexes in photosynthetic processes \cite{Engel2007,Collini2010,Romero2014}.

\subsection*{\centering{4. Conclusions and future work}}

Through this brief overview we have established a number of common principles underpinning  HTSC and biological systems.  In both cases, the interplay between charge density and dissipative and macroscopic quantum forces plays a key role in defining both structure and function.  

 Our understanding of these systems is in its infancy.  Looking to the future, development of a deeper understanding of how these basic principles can be used to develop new materials requires a multidisciplinary program of research.  To date we have considered just a few examples of a much broader range of potential systems dictated by these theoretical principles.  Future work needs to focus on more detailed theory, modelling and controlled experimental studies in both biological and inorganic systems, to determine the impact of atomic/molecular structure, $\rho$, $T$ and other boundary conditions on emergent structures and their properties.  It is clear that we have much to learn from biological systems, such as the use of protein complexes to more precisely control the emergence of structure.  

As a first step, we need to validate theory that structures with higher $D_F$ and $\rho$, do indeed exhibit significantly higher levels of $T_c$.  Confirmation of a clear relationship between $\rho$, $D_F$ and $T_C$ for a specific material would open the door to a new, systematic, {\it{ab initio}} approach to the development of HTSC materials.  At the same time, this work would help confirm theory on macroscopic quantum processes in photosynthetic systems.

\subsection*{Acknowledgements}

We would like to thank H\"ufner {\it{et al}} \cite{Hufner2008} for granting permission to use Fig \ref{SCdome} from their paper. We would also like to acknowledge the role of the COST office, and in particular COST Action FP1105 for facilitating the collaboration that led to this work.

%%%%%%%%%%%%

 %%%%%%%%%%%

%%%%%%%%%%
\end{document}